\documentclass[10pt,a4paper]{natr}
\usepackage[dvips]{graphicx}
\usepackage{wrapfig}
\usepackage{citesort}
\usepackage{helvet}
\begin{document}

\title{The optical spectroscopy of extraterrestrial molecules}
\authors{T.W.~Schmidt $^a$ and R.G.~Sharp $^b$}
\address{$^a$ School of Chemistry, University of Sydney, NSW 2006,
Australia\\$^b$ Anglo-Australian Observatory, PO Box 296 Epping, NSW
1710, Australia} \maketitle
\begin{abstract}
The ongoing quest to identify molecules in the interstellar medium
by their electronic spectra in the visible region is reviewed.
Identification of molecular absorption is described in the context
of the elucidation of the carriers of the unidentified diffuse
interstellar bands while molecular emission is discussed with
reference to the unidentified Red Rectangle bands. The experimental
techniques employed in undertaking studies on the optical
spectroscopy of extraterrestrial molecules are described and
critiqued in the context of their application.
\end{abstract}

\section{INTRODUCTION}

\begin{figure}[h]
\begin{center}
\includegraphics[width = 12 cm]{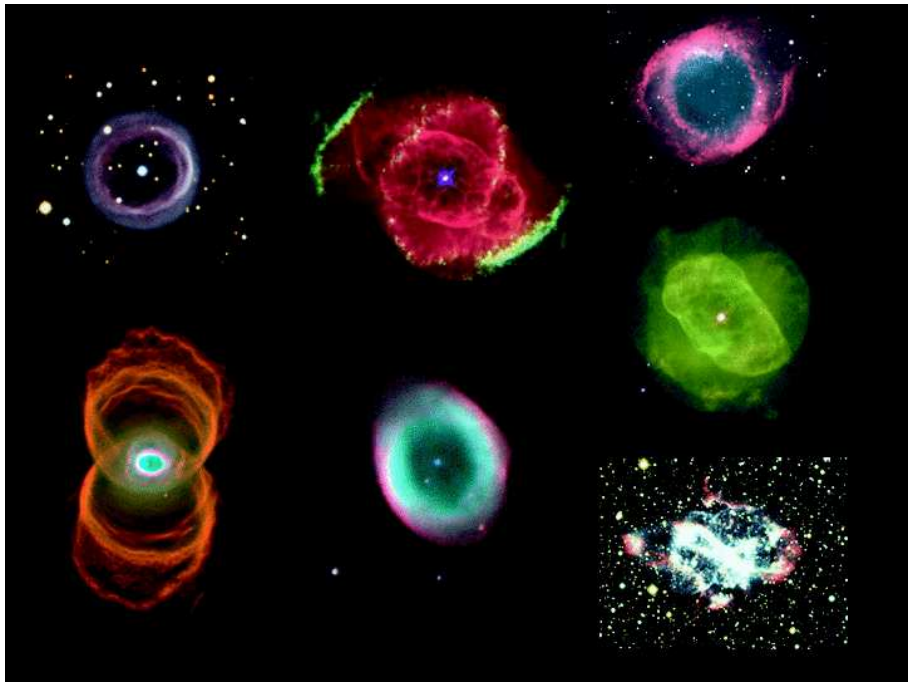}
\end{center}
\caption{\label{nebula} Planetary nebulae as imaged by the Hubble
Space Telescope and the Anglo-Australian Telescope. Planetary
nebulae are gas clouds surrounding stars typically hundreds of light
years away. They take their name from their appearance when imaged
in small telescopes, whereby they resemble gas giant planets from
our solar system such as Uranus or Neptune.  There is no association
between planetary nebular and planets beyond this appearance.}
\end{figure}

As the last embers of a red giant star die down, it undergoes a
series of expansions and contractions, puffing away the outer layers
of the star, resulting in the expulsion of its carbon rich
atmosphere into the cosmos.  As the central stellar core contracts
under gravity into a white dwarf, the atmosphere evolves into a
nascent proto-planetary nebula, rich in carbon, oxygen, nitrogen
(Fig. \ref{nebula}). Such a dignified end to the life of an
intermediate mass star, such as our own Sun is in sharp contrast to
the violent end encountered by higher mass stars \footnote{The mass
limit above which electron degeneracy pressure cannot support a
stellar core against the relentless crushing force of gravity was
first derived by S. Chandrasekhar, for which he was later awarded a
Nobel prize.  Above 1.44 Solar masses a stellar core will collapse
to a neutron star or black hole, resulting in a supernova explosion
rather than the formation of planetary nebular.  Precursor stars
with masses in excess of 1.44 Solar masses may still avoid ending
their lives as supernova if enough mass is lost from the star,
during it's late evolutionary stages, to prevent the remaining core
mass exceeding the Chandrasekhar limit.} which end their lives in
supernova explosions. A supernova is triggered when the collapsing
stellar core is too massive to be supported, against gravitational
collapse, by electron degeneracy pressure (\textit{i.e.} the Pauli
exclusion principle, as is the case in a white dwarf star).  The
resulting violent nuclear explosion heats the surrounding
interstellar medium, through a variety of mechanisms, to
temperatures in excess of $10^6-10^8$\,K, sufficiently hot to
generate X-ray emission. Elements from all over the periodic table
are ejected into the cosmos: the atoms to be later incorporated into
molecular clouds and future solar systems. All elements heavier than
Fe are formed in supernovae.

The chemistry of interstellar space is different to that performed
in a conical flask. It is slow, it is driven by ion-molecule and
neutral-radical reactions in the gas phase \cite{herbst73} and on
the surfaces of dust grains, and it is highly exotic. The harsh
radiation field in interstellar environments is also of great
importance. There are a number of chemical models of interstellar
space. Of note are the ``UMIST gas-phase chemical network'' of
Millar and co-workers in Manchester \cite{umist} and the ``New
Standard Model'' of Herbst and co-workers in Columbus (NSM)
\cite{nsm}. Both these models use complicated networks of kinetic
equations to model the chemistry of various interstellar
environments. However, these models may only be tested by
spectroscopic observation of the relative abundances of interstellar
molecules, which is a field unto itself. Also, the numerous
(thousands of) rate constants are being constantly updated with
better experimental and \textit{ab initio} results.

To understand the chemistry of interstellar clouds one must begin by
first identifying the molecules therein. It is a great challenge
posed by Nature to remotely identify the menagerie of molecules
extant in the interstellar medium (ISM).

Molecules are identified in the interstellar regions by their
spectroscopic signatures in the millimetre, infrared and optical
regions of the electromagnetic spectrum. While it is the millimetre
region which has most greatly illuminated our understanding of the
structures of interstellar molecules, this technique is blind to a
family of molecules of interest: those without permanent dipole
moments. For this reason, the UMIST and NSM models concentrate on
reproducing the observed abundances of polar molecules
\cite{nsm,umist}. An up-to-date list of molecules known to exist in
the interstellar regions is given in Table \ref{list}
\cite{website}.

In the following paragraphs, work concerning the identification of
extraterrestrial molecules in various wavelength regimes is
outlined. The astronomical facilities exemplified are done so in an
Australian context, where possible, and so are not necessarily
indicative of those facilities globally.

\begin{table}
\footnotesize \caption{A list of molecules identified in the
interstellar regions \cite{website}. Carbon chains as long as
HC$_{11}$N have been observed in molecular clouds by millimetre-wave
spectroscopy. Underlined species have been observed due to their
vibration-rotation spectra in the infrared. C$_2$ has only been
observed by electronic spectroscopy (in the optical region).} \vskip
0.3cm
\begin{center}
\begin{tabular}{cl}
\hline\hline No. of atoms & Molecular Formulae\\
\hline 2 & AlF AlCl \textbf{C}$\mathbf{_{2}}$ CH CH$^{+}$ CN CO
CO$^{+}$ CP
CS CSi HCl H$_{2}$ KCl NH NO NS NaCl OH PN SO SO$^{+}$\\
& SiN SiO SiS HF \underline{SH} FeO?\\
3 & \underline{C$_{3}$} C$_{2}$H C$_{2}$O C$_{2}$S CH$_{2}$ HCN HCO HCO$^{+}$ HCS$^{+}$ HOC$^{+}$ H$_{2}$O H$_{2}$S HNC HNO KCN MgCN MgNC \\
& N$_{2}$H$^{+}$ N$_{2}$O NaCN OCS SO$_{2}$ \textsl{c}-SiC$_{2}$ \underline{CO$_{2}$} NH$_{2}$ \underline{H$_{3}^{+}$} AlNC SiCN SiNC H$_2$D$^+$ HD$_2^+$\\
4 & \textsl{c}-C$_{3}$H \textsl{l}-C$_{3}$H C$_{3}$N C$_{3}$O
C$_{3}$S C$_{2}$H$_{2}$ CH$_{2}$D$^{+}$? HCCN HCNH$^{+}$ HNCO HNCS
\underline{CH$_3$} HOCO$^{+}$ H$_{2}$CO \\
& H$_{2}$CN H$_{2}$CS H$_{3}$O$^{+}$
NH$_{3}$ SiC$_{3}$\\
5 & C$_{5}$ C$_{4}$H C$_{4}$Si \textsl{l}-C$_{3}$H$_{2}$
\textsl{c}-C$_{3}$H$_{2}$ CH$_{2}$CN \underline{CH$_{4}$} HC$_{3}$N HC$_{2}$NC HCOOH H$_{2}$CNH H$_{2}$C$_{2}$O H$_{2}$NCN \\
& HNC$_{3}$ \underline{SiH$_{4}$} H$_{2}$COH$^{+}$\\
6 & C$_{5}$H C$_{5}$O C$_{2}$H$_{4}$ CH$_{3}$CN CH$_{3}$NC
CH$_{3}$OH CH$_{3}$SH HC$_{3}$NH$^{+}$ HC$_{2}$CHO HCONH$_{2}$ \underline{\textit{l}-HC$_4$H}?\\
& \textsl{l}-H$_{2}$C$_{4}$ C$_{5}$N \\
7 & C$_{6}$H CH$_{2}$CHCN CH$_{3}$C$_{2}$H HC$_{5}$N CH$_{3}$CHO NH$_{2}$CH$_{3}$ \textsl{c}-C$_{2}$H$_{4}$O  CH$_{2}$CHOH \\
8 & CH$_{3}$C$_{3}$N HCOOCH$_{3}$ CH$_{3}$COOH C$_{7}$H
H$_{2}$C$_{6}$ CH$_{2}$OHCHO \underline{\textit{l}-HC$_6$H}? CH$_2$CHCHO?\\
9 & CH$_{3}$C$_{4}$H CH$_{3}$CH$_{2}$CN (CH$_{3}$)$_{2}$O CH$_{3}$CH$_{2}$OH HC$_{7}$N C$_{8}$H \\
10 & CH$_{3}$C$_{5}$N? (CH$_{3}$)$_{2}$CO NH$_{2}$CH$_{2}$COOH? (CH$_2$OH)$_2$? CH$_3$CH$_2$CHO\\
11 & HC$_{9}$N\\
12 & \underline{C$_6$H$_6$}?\\
13 &HC$_{11}$N\\
 \hline\hline
\label{list}
\end{tabular}
\end{center}
\end{table}

\subsection{The millimetre-wave region}
Molecules are heated by gravitational collapse, converting potential
energy to kinetic energy which is distributed among the degrees of
freedom of the constituent species. This energy can be radiated back
into space by molecules as visible, infrared or millimetre-wave
radiation. Molecules with permanent dipole moments, upon relaxation,
radiate in the millimetre-wave region of the spectrum by cascading
down the ladder of energy levels which describe molecular rotation.
This radiation is collected and analyzed to produce a forest of
sharp, well-defined spectral lines. These lines are matched to
rotational spectra observed in laboratory experiments and in doing
so the extraterrestrial species are identified. Molecules with
larger dipole moments are easier to observe by this technique and as
such asymmetric carbon chains are a dominant motif in the list of
known molecules from the interstellar regions.

Millimetre-wave spectroscopy is performed on extraterrestrial
objects by so-called radio telescopes. Excellent examples of this
type of instrument are the 64\,m Parkes radio telescope, the 22\,m
Mopra telescope, and the Australia Telescope Compact Array, all
administered by the Australia Telescope National Facility of the
CSIRO \cite{atnf} (Fig. \ref{astro}). There are many groups who
undertake laboratory experiments to which astronomical observations
may be compared. In the laboratory, rotational spectroscopy is
performed by a fourier-transform technique whereby the rotational
spectrum is obtained in a similar fashion to the free-induction
decay well known in the field of nuclear magnetic resonance. The
group of Thaddeus and co-workers have discovered over fifty
molecules of astrophysical relevance \cite{thaddeus01}.

\subsection{The infrared region}
The infrared region features emission corresponding to vibrational
relaxation of specific functional groups and bonds comprising
interstellar molecules. Of note is the 3.3\,$\mu$m emission lines
which are thought to originate from polycyclic aromatic hydrocarbons
(PAHs), such as naphthalene, anthracene and phenanthrene.
Spectroscopy in the infrared is difficult, however, due to the
interference of sources of background infrared emission (sky and
telescope). Some regions of the infrared are ``off-limits'' to
astronomers due to absorption of radiation by water in the
atmosphere. Emission in the far infrared has been used to identify
interstellar C$_3$ \cite{hinkle88} and C$_5$ \cite{bernath89}.

\begin{figure}[h]
\begin{center}
\includegraphics[width = 3.5 cm]{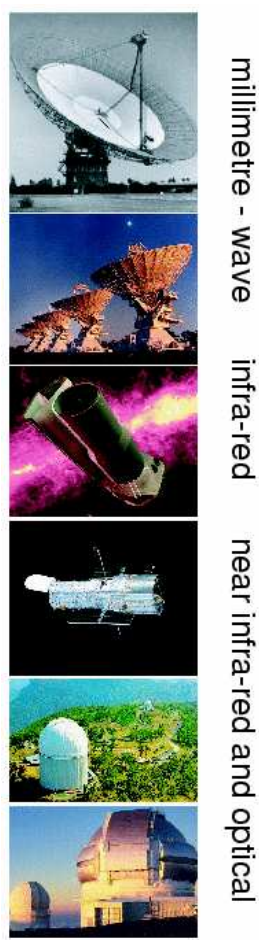}
\caption{Astronomical instruments used to observe extraterrestrial
molecules. Top to bottom: Parkes radio-telescope; Australia
Telescope Compact Array; Spitzer; Hubble Space Telescope; AAT at
Coonabarabran; Gemini North (CFH in background).\label{astro} }
\end{center}
\end{figure}

Due to technical difficulties associated with ground-based
observations in the infrared, last year (2003), a new satellite was
launched by NASA. This new instrument, SPITZER, (Fig. \ref{astro})
is capable of performing infrared spectroscopy in the wavelength
region spanning 3-180\,$\mu$m at various levels of resolution. The
ground-based Michelle (Mid-Infrared Echelle spectrograph) mounted on
Gemini North, Hawaii, is capable of spectroscopy in the 7-26\,$\mu$m
region with resolution of 30000
\footnote{$R=\lambda/\Delta\lambda$}. The Anglo-Australian Telescope
(AAT) at Coonabarabran is capable of spectroscopy in the
0.9-2.5\,$\mu$m range with resolution of 2400 \cite{aat}.

In the laboratory, infrared spectroscopy is routinely performed
using a fourier-transform technique (FTIR). However, this technique
is less sensitive than tunable diode laser spectroscopy (TDL). TDL
spectroscopy as an absorption technique has been applied to the IR
spectroscopy of many carbon chains and rings of astrophysical
relevance. Of particular note is the Cologne Carbon Cluster
experiment \cite{koln} of Winnewisser and co-workers.

\subsection{The optical region}
It is in the optical regions of the electromagnetic spectrum
(300-950\,nm) that there is much to be done. Despite the optical
region being the part of the electromagnetic spectrum originally
accessed by astronomers, there have been scarce new identifications
of interstellar molecules by their electronic spectra. Examples of
astronomical instruments available in Australia that access this
region for the purposes of optical spectroscopy are the ultra-high
resolution facility (UHRF) ($R\sim900,000$) and the University
College London Coud\'e Echelle Spectrograph (UCLES) on the AAT
\cite{aat}. Facilities with Australian access include the Gemini
Multi-Object Spectrograph (GMOS) on Gemini North (Hawaii) and South
(Chile), which is a low resolution optical spectrograph. The
Bench-mounted High Resolution Optical Spectrograph (bHROS) on Gemini
South will be in routine operation shortly.

Molecules without permanent dipole moments may not be observed by
millimetre wave spectroscopy, and those without infrared active
vibrational modes cannot be observed by spectroscopy in that region.
For instance, C$_2$ may only be observed by optical spectroscopy.
For these reasons, spectroscopy in the optical region offers
possibilities for observing new interstellar molecules. This review
focusses on the research being undertaken in the field of the
identification of \emph{new} interstellar molecules through their
electronic spectra in the near-infrared, visible and ultraviolet
spectrum.

The article is arranged as follows. Firstly, the absorption of
starlight by molecules is discussed and astronomical and laboratory
work in this area is reviewed in the context of the unidentified
``Diffuse Interstellar Bands'' (DIBs). Molecular emission is
discussed in relation to the Red Rectangle paradigm and ongoing work
in this area. The experimental techniques and results involved in
performing interstellar spectroscopy in the laboratory are reviewed
in the context of their application. Finally, state-of-the-art
experiments are described and suggestions are made as to the future
directions of the field.

\section{Molecular absorption - the \textsl{Diffuse Interstellar Bands}}
The spectrum of light from many stars is well described to first
order by black-body radiation theory: the hotter the star, the bluer
its spectrum. So-called ``reddened'' stars are those for which the
blue part of the spectrum is attenuated by scattering. The
scattering takes place in the interstellar clouds, along the line of
sight to the background star.  The passage of the starlight though
such a cloud can leave the signature of the DIBs imprinted on the
stellar spectrum. With starlight as a ``white-light'' source, the
interstellar cloud as a sample, and with spectrographs on Earth, one
has all the essential elements of a benchtop spectrophotometer
(albeit slightly larger).

Molecular clouds vary in their density. The denser clouds greatly
diminish the brightness of the background star thereby also
sheltering the interior of the cloud from harmful deep ultraviolet
radiation. The denser clouds naturally have higher abundances of
molecular species, which are measured by ``column density''. Column
density is an effective number of molecules in a column of space
between the observer and the light source, typically quoted as
having a 1\,cm$^2$ cross-section.

The spectra of many diatomic molecules have been known for decades
\cite{huber}. Diatomic species, such as CH, C$_2$, CN, and CH$^+$,
have been detected toward a number of clouds. Their column densities
are regarded as standards with which chemical models of interstellar
clouds may be compared. Less is known about absorption in the
optical region by polyatomic molecules.

In the spectra taken towards diffuse clouds, there are approximately
300 absorption features of unknown origin. These features,
collectively known as the ``Diffuse Interstellar Bands'' (DIBs),
vary in width from 0.1-3\,nm and cover the entire visible and near
IR regions. There is a vast body of phenomenology pertaining to the
identity of the DIB carriers, which has been reviewed elsewhere
\cite{herbig}. There are various hypotheses as to which types of
transitions are responsible for the DIBs \cite{fulara00}. These are
outlined below.

\subsection{The carbon chain hypothesis}
Carbon chains were first proposed as the carriers of the DIBs by
Douglas \cite{douglas77}. That the DIB absorbers are carbon chains
is predicated upon the observation that carbon chains can exhibit
very strong transitions in the visible region, and that they are
known to exist in molecular clouds \cite{fulara93a}. The list in
Table \ref{list} demonstrates that carbon chains are a widely
exhibited motif of the known interstellar molecules. The carbon
chains observed by millimetre-wave astronomy are necessarily
strongly polar (\textit{e.g.} $\mathrm {H-C\equiv C-C\equiv
C-C\equiv C-C\equiv C-C\equiv C-C\equiv N}$) yet it is expected that
the bare carbon chains (C$_n$) will also be present in molecular
clouds. It is hypothesised that some or all of the DIB absorbers are
carbon chain molecules.

The hypothesis is easy to prove, at least in principle. One must
measure the absorption spectra of target molecules, in the gas
phase, under isolated conditions, and compare these to astronomical
spectra. An example of a positive identification of an interstellar
carbon chain by this method is the observation of C$_3$ in
interstellar clouds \cite{maier01}. C$_3$ had been previously
identified in the laboratory and its spectrum at 405\,nm was well
known from laser-induced fluorescence spectroscopy \cite{rohlfing89}
and comets \cite{comets}.

\begin{figure}[h]
\begin{center}
\includegraphics[width = 15 cm]{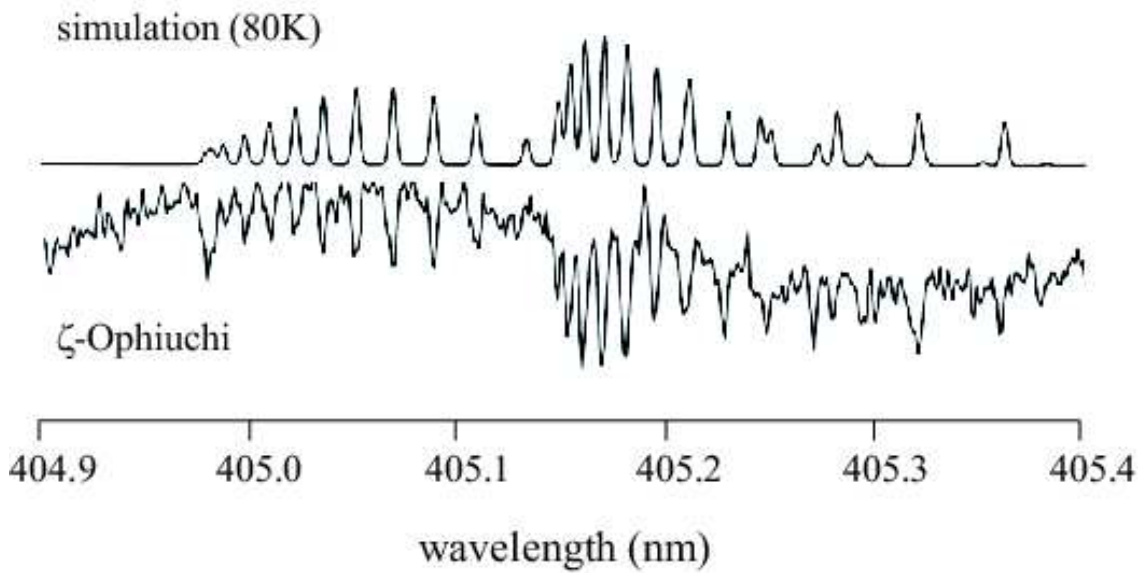}
\end{center}
\caption{\label{c3} The absorption spectrum of C$_3$ observed by
Maier \textit{et al} towards $\zeta$-Ophiuchi as compared to a
simulation from known line positions at 80\,K. (Adapted from Ref.
\cite{maier01})}
\end{figure}

A survey towards $\zeta$-Ophiuchi, 20-Aquilae and $\zeta$-Persei
revealed absorption by the $A^1\Pi_u\leftarrow X^1\Sigma_g^+$
transition of C$_3$ \cite{maier01}. The rotational profile was
fitted to 80\,K and the column density was determined to be
$1-2\times10^{12}$\,cm$^{-2}$ (Fig. \ref{c3}). Since then, C$_3$ has
been observed towards many stars with column densities reported up
to $10^{13}$\,cm$^{-2}$. A strong relationship between the column
densities of C$_2$ and C$_3$ with
$N(\mathrm{C}_2)/N(\mathrm{C}_3)\simeq40$ has been reported
\cite{oka03}. So far, searches for C$_4$ and C$_5$ by optical
spectroscopy have been unsuccessful \cite{maier04}. Upper bounds
have been placed on column densities of C$_4$ and C$_5$ towards
$\zeta$-Ophiuchi of $10^{13}$\,cm$^{-2}$ and
2$\times10^{11}$\,cm$^{-2}$ respectively. The value of
\textit{N}(C$_4$) relies on a calculated oscillator strength and is
therefore less certain. It should be pointed out that C$_5$ and
C$_3$ have been observed by high resolution IR absorption
spectroscopy in carbon-rich nebulae \cite{hinkle88,bernath89}.
Carbon chains, bare, monohydrogenated or dihydrogenated are expected
or have been shown to exist in the ISM. Whether they are the DIB
absorbers can be rigorously tested by a combination of laboratory
spectroscopy and observation. A large part of the work reviewed here
was performed in Basel, Switzerland, in the group of J.P. Maier.

\subsubsection{Resonant 2-Colour 2-Photon Ionization (R2C2PI) spectroscopy}
\begin{figure}[h]
\begin{center}
\includegraphics[width = 12 cm]{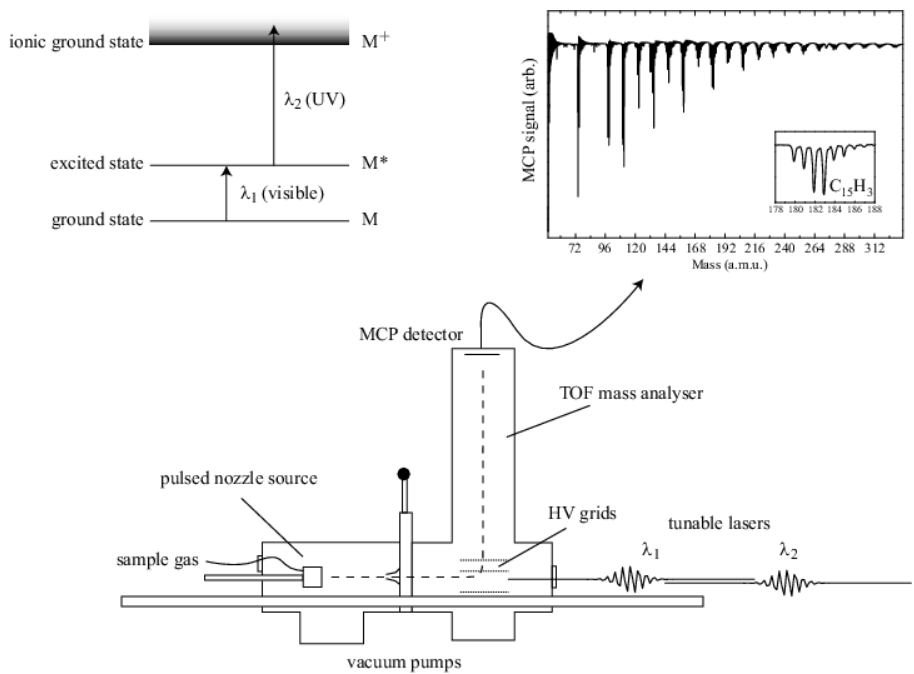}
\end{center}
\caption{\label{r2c2pi} In R2C2PI spectroscopy, supersonically
cooled molecules are ionized in two steps by photons of two colours
(top-left) and then mass-analyzed. The energy of $\lambda_2$ is not
energetic enough to ionize the ground state molecule in a one photon
process, so appreciable ion signal is only observed when $\lambda_1$
is resonant with an excited state of the molecule. As the wavelength
$\lambda_1$ is scanned, an excitation spectrum is produced. The mass
spectrum of a hydrocarbon discharge appears as indicated in the
top-right of the figure. When $\lambda_1$ is resonant, a mass peak
will be enhanced by orders of magnitude in strength.}
\end{figure}

The spectra of dihydrogenated chains, HC$_n$H, were first observed
in solution and in solid matrices. The gas-phase spectra of the even
series, HC$_{2n}$H, were recorded only recently. The spectra of
HC$_{2n}$H ($n=8-13$) were obtained by Resonant 2-Colour 2-Photon
Ionization (R2C2PI) spectroscopy in a molecular beam produced by a
discharge of diacetylene in argon \cite{pino01}. R2C2PI spectroscopy
is a type of Resonance-Enhanced Multi-Photon Ionization (REMPI)
spectroscopy \cite{boesl91} (see Fig. \ref{r2c2pi}). In such an
experiment, the gas-phase supersonically cooled products of the
discharge (containing C$_n$H$_m$ species with $n\geq m$) are
irradiated by two laser beams. The first is scanned in wavelength.
If a photon from the first laser is absorbed, then absorption from
the second laser will overcome the ionization potential of the
molecule and nascent ions will be produced. The ions are accelerated
towards a detector, arriving at a time-of-flight (TOF)
characteristic of the mass to charge ratio of the ion (generally
speaking only singly charged molecules are observed). This
experiment yielded the spectra of the even chains up to HC$_{26}$H,
none of which were observed to lie in the visible region (the
longest chain, HC$_{26}$H, absorbs at 340\,nm). The reason for this
behaviour is bond length alternation. The hydrogen end-caps of the
chain induce triple-bond/single-bond alternation which slowly decays
toward the molecular centre. The overall effect is that of the chain
exhibiting a bandgap, which slows the movement of the absorption
positions to lower energies as the chain length is increased
\cite{ding03}. Very weak, forbidden, bands of the HC$_{2n}$H series
have been observed in the visible region, yet these are not of
relevance to astrophysical studies due to their low oscillator
strengths \cite{ding03}.

\subsubsection{Cavity Ringdown Spectroscopy (CRDS)}
\begin{figure}[h]
\begin{center}
\includegraphics[width = 12 cm]{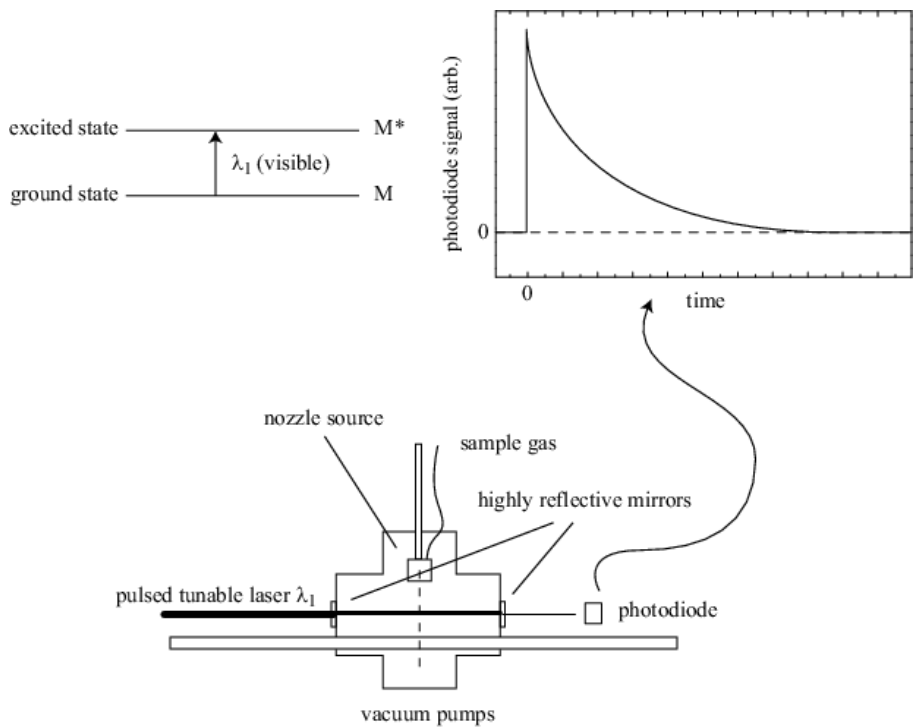}
\end{center}
\caption{\label{crds} In cavity ringdown spectroscopy (CRDS),
supersonically cooled molecules are injected into an optical cavity
in a vacuum chamber. The decay profile of the laser pulse as
observed by a photodetector exterior to the cavity is modulated by
absorption by molecular species. The modulation of the decay profile
as a function of laser wavelength yields CRDS spectra.}
\end{figure}

Absorption spectra of the odd chains HC$_{2n+1}$H ($n=3-6$) were
observed in the visible region by cavity ringdown spectroscopy
(CRDS) \cite{ball00}. CRDS can be performed using pulsed lasers
\cite{scherer97,okeefe88} or continuous wave lasers \cite{birza02}.
The principle is as follows (see Fig. \ref{crds}). molecules are
expanded into a vacuum chamber such that they cool supersonically.
In experiments related to carbon chains, the molecules expanded into
the vacuum are the products of a hydrocarbon discharge, much like
the R2C2PI experiments outlined above \cite{birza02}. The free jet
expands in an optical cavity defined by two highly reflective
mirrors mounted on either side of the vacuum chamber. A laser beam
is injected into the cavity (tuned to support the particular
wavelength). The decay of the light pulse in the cavity, as observed
by a photodetector mounted exterior to the vacuum chamber, can be
related to the reflectivity of the mirrors, and absorption of laser
light by some molecular species extant in the free jet expansion.
Scanning the laser beam produces spectra which must be analysed by
the rotational structure, for there is no mass information about the
absorber (unlike R2C2PI spectroscopy). Nevertheless, the spectra of
HC$_{2n+1}$H ($n=3-6$) in the visible region were recorded in this
way \cite{ball00}. These spectra were not found to match any DIBs.

A simple free electron model of the electronic structure showed that
the oscillator strength for these visible transitions increased with
chain length. High-level MRCI calculations and CASSCF calculations
did not support this assertion
\cite{muhlhauser02,mpourmakis02,ding03b}. It was found by
computation that the excited states of the HC$_{2n+1}$H series could
be described by an admixture of two determinants. The determinants
combined in even and odd combination to produce a lower energy
excited state which carried very little oscillator strength and a
higher energy state which carried a very large oscillator strength.
The lower energy $\tilde{A}$ states, represented by absorption of
HC$_{2n+1}$H ($n=3-6$) in the visible region were thus found to be
irrelevant to the DIB problem, on account of their vanishing
oscillator strengths. This realization spurred a search for the
$\tilde{B}$ states by R2C2PI spectroscopy. Two bands were observed,
for HC$_{13}$H and HC$_{19}$H, both lying in the UV region
\cite{ding03b}. It is concluded that dihydrogenated chains cannot be
responsible for DIBs unless they are considerably longer than
HC$_{19}$H (it is estimated that HC$_{30}$H will absorb in the
visible region).

Due to their strong dipole moment, the monohydrogenated chains,
C$_{2n}$H, have been observed in the ISM by rotational spectroscopy
\cite{thaddeus01}. The optical spectra of monohydrogenated chains
have been observed in the visible region by CRDS and R2C2PI
spectroscopy \cite{ding02,linnartz98}. That the spectra did not
match any known DIBs places upper limits on their column densities
of $\approx10^{12}$\,cm$^{-2}$. The optical transitions observed for
the C$_{2n}$H chains do not have large oscillator strengths. In
general, polarization of the carbon chain reduces the overlap
between the highest occupied molecular orbital and the lowest
unoccupied molecular orbital, thereby reducing the possible
oscillator strength of the transition.

Of the bare carbon chains, C$_3$ and C$_5$ have been observed in the
interstellar medium \cite{hinkle88,bernath89,maier01}. The
absorption by C$_3$ towards $\zeta$-Oph is due to a relatively weak
perpendicular transition, first observed in fluorescence in comet
tails. There is a much stronger parallel transition of C$_3$ in the
vacuum ultraviolet region \cite{monninger02}. This strong parallel
transition is seen analogously in all C$_{2n+1}$ chains. The
oscillator strength grows with the size of the chain and the
wavelength shifts linearly (as opposed to HC$_{2n}$H which exhibits
a band gap). The absorption spectra of C$_{7}$ to C$_{21}$ have been
observed in a neon matrix \cite{maier98,wyss99}. Absorptions due to
the strong parallel transition all lie in the visible region yet gas
phase spectra, with which astronomical observations may be compared,
have so far proved elusive. It is the authors' opinion that a carbon
cluster source coupled to the R2C2PI technique represents the best
chance of observing these spectra in the laboratory.

Gas-phase spectra of the even carbon chains have also proved
elusive. Condensed phase spectra of several members of the C$_{2n}$
series have been observed \cite{rechtsteiner01,grutter99}. The
transitions observed often lie in the visible region and thus these
molecules cannot be ruled out as the DIB carriers. Definitive proof,
one way or the other, will come with unambiguous gas-phase spectra.

Open shell carbon chains are not the best candidates as DIB
absorbers, as their transitions are highly mixed, either
distributing oscillator strength across many transitions or shifting
the oscillator strength into the UV for medium length chains. Of the
closed shell chains, there are HC$_{2n}$H, HC$_{2n+1}^+$, and
C$_{2n+1}$. As noted above, the polyyne series, HC$_{2n}$H, suffers
from a non-linear dependence of absorption wavelength with chain
size. The gas-phase spectrum of HC$_{26}$H lies in the near UV and
only much longer chains will begin to absorb in the visible
($2n\sim40$). Of the HC$_{2n+1}^+$ cations, not much is known. Their
spectra should mimic the C$_{2n+1}$ neutrals however the transitions
will not be quite so strong since the overlap between the highest
occupied and the lowest unoccupied molecular orbital is reduced in
the polar molecules.

\subsubsection{Photodetachment spectroscopy} 
\begin{figure}[h]
\begin{center}
\includegraphics[width = 12 cm]{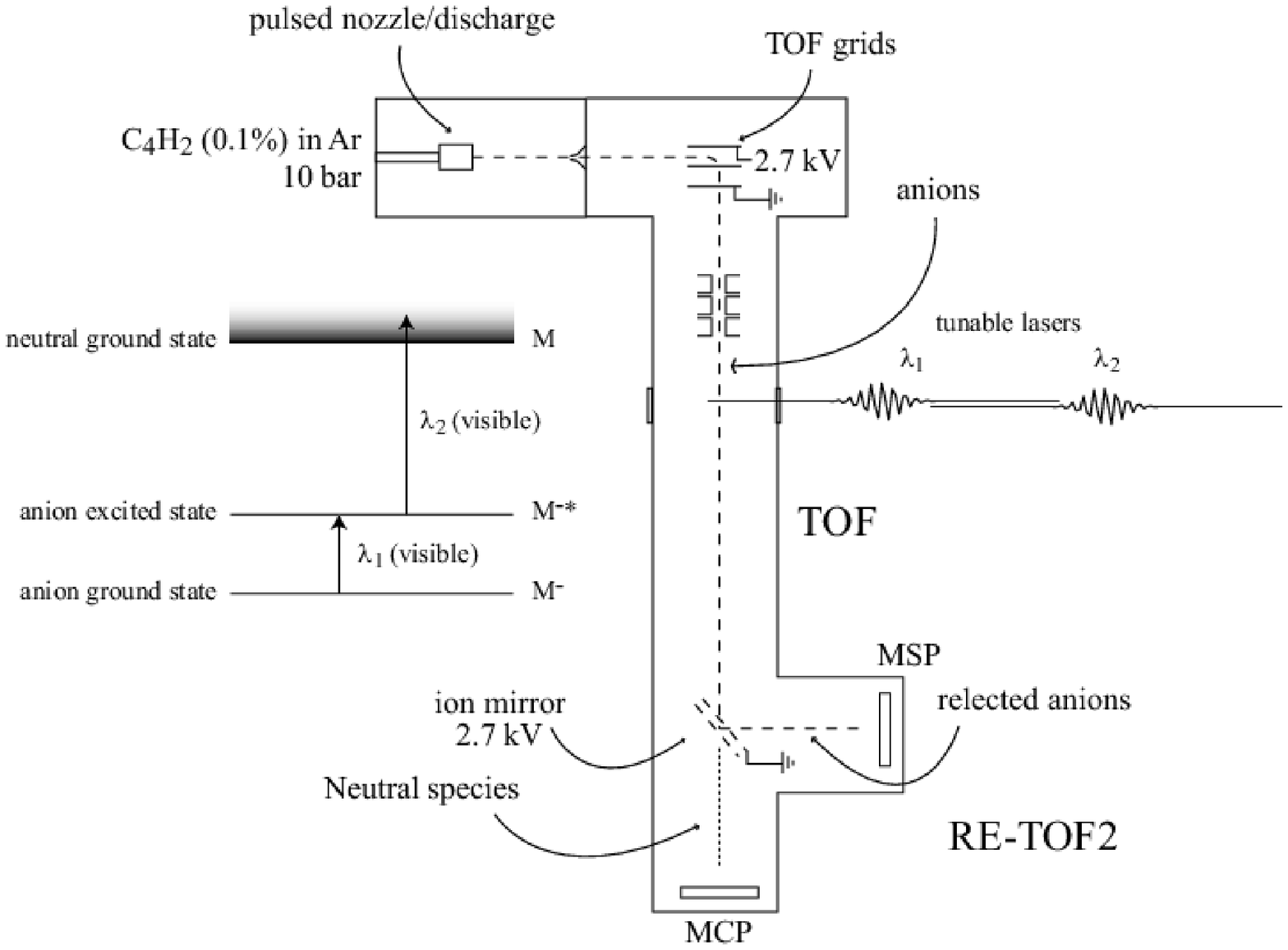}
\end{center}
\caption{\label{anions} An illustration of the apparatus used to
produce the spectra of carbon chain anions. Anions are produced in a
discharge and then accelerated in a time-of-flight (TOF) tube. The
mass-selected ion bunch is intercepted by a pair of laser pulses
which coorperatively neutralize the anion. The neutral signal as a
function of $\lambda _1$ is the photodetachment spectrum.}
\end{figure}

It would be remiss not to discuss the possibility of
carbon chain anions in the ISM. Indeed, carbon chains have been
found to possess very high electron affinities \cite{tulej00}. In
particular, polar carbon chains may be efficient at electron capture
due to the existence of dipole-bound states \cite{sarre00,guethe01}.
The spectrum of C$_7^-$ was observed to match very closely several
DIBs \cite{tulej98,lakin00}. More precise observation showed that
the match was not exact \cite{mccall00}, and as such C$_7^-$ was not
responsible for any DIBs. The spectra of anions are recorded in the
gas phase is a similar manner to R2C2PI spectra. The set-up is
illustrated in Fig. \ref{anions}. Anions are produced in a
hydrocarbon discharge and accelerated in a time of flight (TOF) tube
to the laser interaction region. The mass-selected anion bunch is
intercepted by a pair of laser beams which cooperatively excite the
anion and subsequently photodetach an electron. The nascent neutral
molecules are oblivious to the ion mirror which reflects the
remaining anions. The neutrals impacting onto the MCP at 2.7\,keV
induce a signal due to production of secondary electrons. The
neutral signal as a function of the first laser pulse yields the
excitation spectrum of the anion.

While the possibilities are numerous, the outstanding candidates as
carbon chain carriers of the DIBs are the odd-numbered carbon
chains. It is expected that this hypothesis will be tested within
the next few years. The examples of carbon chains terminated with
heteratoms \cite{denisov04,ding04} are too numerous to review here.
However, they have been observed in molecular clouds (see Table
\ref{list}) and are likely important species in interstellar
chemistry \cite{petrie03}. In time, some of these will be detected
in the interstellar medium by optical spectroscopy.

\subsection{The PAH hypothesis}

\begin{figure}[h]
\begin{center}
\includegraphics[width = 12 cm]{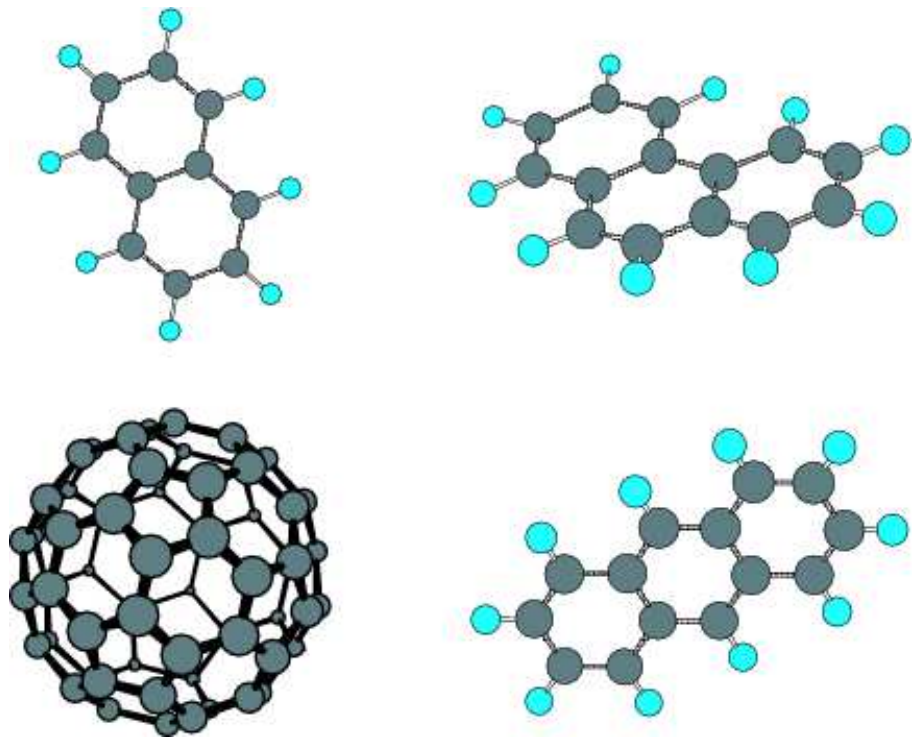}
\end{center}
\caption{\label{pah} Structures of cations thought to exist in the
interstella medium. clockwise from top-left: naphthalenylium cation,
phenanthrenylium cation, anthracenylium cation and
buckminsterfullerenylium cation (C$_{60}^+$).}
\end{figure}

Polycyclic aromatic hydrocarbons (PAHS) are a class of molecule
characterized by conjoined ``benzene ring'' moieties. They may be
thought of as fragments of graphite with hydrogens bound to the
edge. Examples of PAHs are naphthalene, anthracene (see Fig.
\ref{pah}).

The presence of PAHs in the ISM was first proposed in the context of
their possible role in the form of the UV extinction curve
\cite{donn68}. The suggestion that PAHs might be the carriers of the
DIBs came later \cite{vanderzwet85,leger85}, with the realization
that in order that smaller PAHs absorb in the visible they should be
in cationic form \cite{crawford85}.

With the ``coming of age'' of Mid-IR astronomy, particularly results
from the European Space Agency (ESA) Infrared Space Observatory
(ISO) mission such as the identification of interstellar benzene
\cite{cernicharo01}, PAH molecules have been interpreted as the
natural carriers for the observed Mid-IR emission band
\cite{allamandola89}.  Indeed, recent work has seen the first
indications that the Mid-IR emission features can be well fitted by
composites of emission from numerous PAH species
\cite{allamandola99}. Unfortunately, the composite nature of the
Mid-IR bands makes this region of the spectrum poor for
identification of specific PAH molecules. The promising
possibilities of using Far-IR emission bands (vibrational
frequencies associated with the bending of the skeletal structure of
molecules) to uniquely identify PAH molecules has been discussed
\cite{mulas03}.

Further support for the PAH origin of DIBs comes from their spectral
stability (with respect to environmental variations in the
excitation spectrum).  While the Mid-IR PAH features are observed to
vary greatly in structure and intensity ratios between observations,
the DIBs are observed to be surprisingly uniform in structure, with
only the relative intensity of different bands altering with sight
line, presumably an effect of differing abundances of species along
different sight line.  It has been demonstrated that environmental
effects are significant in altering the emission profiles of the
Mid-IR PAH features \cite{allamandola99}.  This can account for the
large variations seen in the observed Mid-IR emission and the
discrepant intensity ratios when compared with laboratory
observations. However, it has been shown that this is not the case
for visible light PAH transitions \cite{malloci03}.  They show that
PAH absorption features would be observed to be stable, at the level
of current observations, over an extreme range of environmental
conditions.

Circumstantial evidence in favor of larger molecular carbon species
is reported \cite{cardelli96} whereby a potential ``carbon-crisis''
is highlighted: insufficient carbon is identified in ISM to support
the proposed build up of carbon rich dust grains. High molecular
weight PAH molecules could act as repositories for carbon. It has
been estimated that up to 20\% of cosmic carbon in the Galaxy is
contained in PAHs \cite{dwek97} yet more likely it is less than
this. The current status of possible matches between DIBs and
laboratory PAH features has been reviewed \cite{salama99}. With the
presence of PAH molecules in the ISM now widely accepted, the search
for transitions in the visible region which could give rise to the
DIBs is a logical next step. This is a problem for laboratory
astrochemistry.

Since most neutral PAHs amenable to spectroscopic study absorb only
in the UV, research has concentrated on measuring the spectra of the
PAH cations in the visible region. Until recently, only matrix
isolation spectra of PAH cations were available. The first gas phase
spectrum of a PAH cation was measured by resonance enhanced
dissociation spectroscopy of naphthalene cation, Np$^+$
\cite{syage87}. The spectrum produced was noisy and not definitive.
It was not until 1999 that the absorption spectrum of the
naphthalene cation (the smallest PAH) was recorded
\cite{romanini99,pino99}. Two bands were recorded by CRDS
(\textit{vide supra}) in a pulsed discharge source. Following this,
the entire spectrum of the Np$^+$ cation was produced using an
action spectroscopy technique pioneered earlier in work on benzene
analogs \cite{mckay90}. Np$^+$ was clustered with argon ``spectator
atoms''. The cluster mass distribution as measured by time-of-flight
mass spectrometry was observed to change when energy was absorbed by
the cationic chromophore. Essentially, the spectator atom is
evaporated when the Np$^+$ chromophore absorbs a visible photon, and
the mass change is recorded. The ``action'' as a function of
wavelength yields a proxy excitation spectrum. This technique is
powerful however suffers from the spectator atoms inducing slight
changes in the electronic chromophore. The change (0.1\,nm at
648\,nm) is enough to exclude these spectra from being directly
comparable to astronomical observation. More recently, acephenene
and naphthalene cations have been studied in more detail by CRDS
\cite{ludovic03}. The technique which yielded the first PAH
excitation spectrum, namely resonance-enhanced multiphoton
dissociation (REMPD) has undergone a small revival. The development
of ion-traps had led to the ability to irradiate a population of
ions with multiple photons, thereby accumulating signal before
detection. Mass-selected ions are trapped in the gas phase and
irradiated with several laser shots. Mass spectrometry of the
fragments reveals acetylene loss as an indicator of the ions having
absorbed energy. The probability of the ion having absorbed enough
to break carbon-carbon bonds is greatly enhanced if the first photon
may be absorbed in such a way as to place the ion in an excited
state. In this way, warm gas phase spectra of naphthalene and
anthracene have been recorded \cite{rolland03}. Mounting an ion trap
on a cold head is one way of alleviating the problems of temperature
on the spectrum \cite{schlemmer02}.

At the time of writing, not one PAH, neutral or cationic, has been
identified in the ISM by optical spectroscopy. Indeed, not one DIB
carrier has been positively identified. While it is certain that
PAHs exist in the ISM, it is unclear whether they should be
dominantly ionized or neutral. As PAHs are likely to have large
electron affinities, there is also scope for the existence of PAH
anions in the ISM. Indeed, the electron affinities of some
carbon-based molecules exceed the ionization energy of alkaline
earth atoms. As such, in an environment where one finds sodium
atoms, one may also find carbonaceous anions. It is also unclear as
to whether the PAHs should be wholly intact. One conclusion that may
be drawn from the observed PAH cation spectra is that they are most
probably not responsible for narrow DIBs, because the PAH cation
bands observed thus far are all very broad ($\approx10\AA$),
presumably due to lifetime effects associated with internal
conversion processes. The Possibilities regarding extensions of the
PAH hypothesis will be discussed below.

\subsection{Something completely different?}
There have been many suggestions over the years as to the identity
of the DIB absorbers. Those suggestions taken most seriously, namely
carbon chains and PAHs, have been discussed above. The remaining
candidates are plentiful, and only need be tested in the laboratory.

The possible existence of buckminsterfullerenes in space was first
suggested by Kroto \cite{kroto89}. That C$_{60}$ in particular is so
symmetrical and so stable lends credence to the hypothesis. However,
the strongest absorptions of C$_{60}$ occur in the UV and as such
this molecule is not responsible for any DIBs \cite{frum91}.

Following the publication of the matrix isolation spectrum of the
C$_{60}^+$ cation \cite{fulara93b}, a search was carried out towards
a number of stars which revealed two new DIBs. The DIBs were found
to have a spacing and absorption wavelength consistent with the
observed spectrum of C$_{60}^+$ in an argon matrix. While the
assignment of the DIBs observed near 9500\AA\ to C$_{60}^+$ seems
entirely reasonable, definitive proof can only come by the
laboratory gas phase spectrum of C$_{60}^+$, which has so far proved
elusive. C$_{60}^{2+}$ may also exist in appreciable concentrations
in the ISM, since the second ionization potential of C$_{60}$
 is extraordinarily low (11.4\,eV) \cite{steger92}.

The column density of H$_2$ in molecular clouds is approximately
$10^6$ times greater than that of the most abundant polyatomic
carbon species. This implies that for a given oscillator strength of
a transition, only one millionth of the H$_2$ present need be in a
particular state to effect the same absorption as other proposed
carriers of the DIBs. Indeed, there exist inter-Rydberg transitions
of H$_2$ calculated to match DIBs very well
\cite{sorokin96,ubachs97}. Experimental observations of the
inter-Rydberg transitions have revealed intriguing properties of
H$_2$ such as ``outer well'' states with W-shaped potential energy
curves \cite{reinhold98,koelemeij03}. Another theory put forward
attributes the DIBs to Rydberg matter \cite{holmlid04}, aggregations
of excited atoms and molecules bonded through electrostatic
interactions of Rydberg electrons.

The existence of interstellar diamonds was suggested in 1969
\cite{saslaw69}. They have since been shown to exist as nanometre
sized crystallites in carbonaceous meteorites \cite{lewis87}. Given
the ability for defects and surface effects to produce colour
centres, nanodiamonds present themselves as possible candidates as
the carriers of some of the DIBs.

\begin{figure}[h]
\begin{center}
\includegraphics[width = 6 cm]{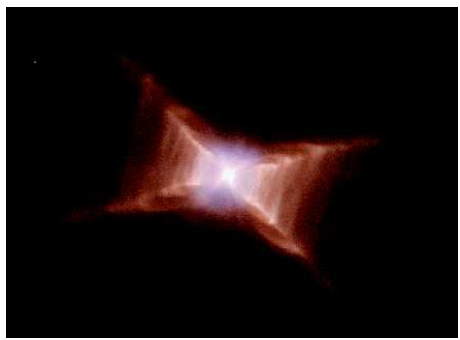}
\end{center}
\caption{\label{rr} The ``Red Rectangle'', a nearby proto-planetary
nebula, is a carbon-rich object in which there are unidentified
emitters, thought to be molecular \cite{hst} (Image reproduced with
permission of the authors of reference \cite{hst}.)}
\end{figure}

\section{Molecular emission - The Red Rectangle}

The search for molecules in Space by optical spectroscopy may be
performed by the molecules' absorption or emission of visible and UV
light. The search for molecules by absorption is intrinsically
linked to the search for the DIB carriers. In the case of C$_3$, the
molecule was first observed by emission in comet tails, subsequently
in the laboratory and then finally by absorption in interstellar
clouds \cite{maier01}.

The ``Red Rectangle'' (see Fig \ref{rr}) is a biconical
proto-planetary nebula. In the case of the Red Rectangle, the core
is a binary system. One of the stars has come to the end of its life
and has started puffing off its atmosphere to leave behind a white
dwarf. The other component is a helium white dwarf which died
previously in a similar manner. It will have been more massive,
leading to an earlier evolution. Much of its mass will most likely
have been accreted onto the other star, rather than puffed off in a
previous nebula. The system will evolve into a binary white dwarf
system.

\begin{figure}[h]
\begin{center}
\includegraphics[width = 15 cm]{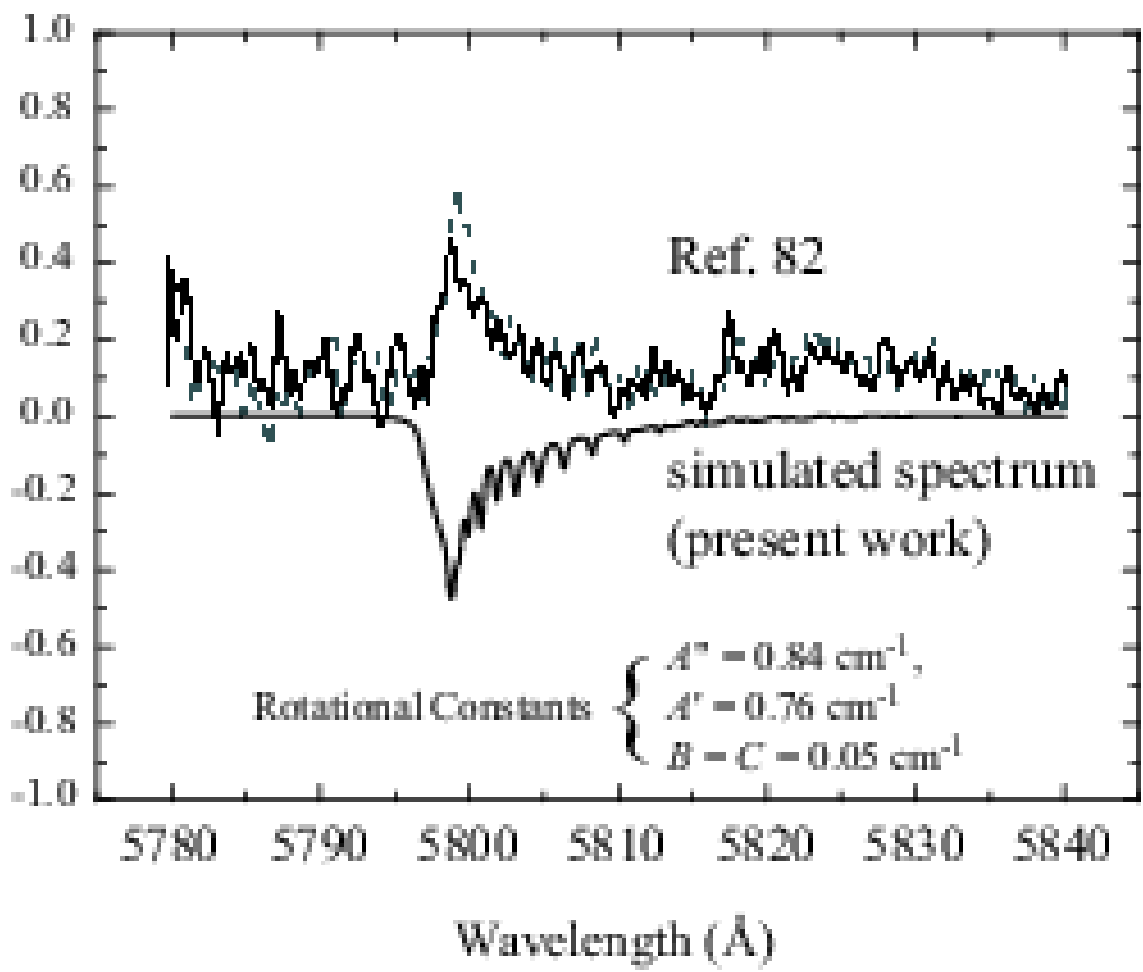}
\end{center}
\caption{\label{rrspec} A high resolution portion of the extended
red emission of the Red Rectangle as compared to a simulation.}
\end{figure}

The unusual geometry of the nebula is not entirely understood
\cite{menshchikov02}. The Red Rectangle exhibits a ``bipolar flow''
which carries mass away from the central stars into the interstellar
medium. It is suggested that the central stars give rise to a pair
of jets that precess about one another (like a spinning top). The
nebula's emission is also unusual: it displays a number of
unidentified emission lines, the so-called Red Rectangle Bands
(RRBs) \cite{glinski02}. These bands, occuring in the visible region
are speculated to be due to unusual carbon containing molecules.
There is also suggestion that at least one of the features is
related to a DIB \cite{jenniskens96}. Thus, identification of the
carrier of the band may be a ``foot in the door'' to the
identification of the DIB carriers. This feature is illustrated in
Fig. \ref{rrspec}.

The advantage of looking for molecules in emission is that the
species necessarily fluoresce, or phosphoresce. In the case of
fluorescence, the carriers may be observed by laser induced
fluorescence (LIF) spectroscopy, so long as the species can be
created in the laboratory. One possible problem is that the LIF
spectrum of a hydrocarbon discharge may be too rich to positively
identify individual species. However, there has been much progress
in the last five years in diagnosing the products of a hydrocarbon
discharge by gas-phase spectroscopy. It is now possible that much of
the fluorescence can be assigned. The remainder will belong to new
molecules. Identification of mass-unresolved spectra will come about
by a combination of \textit{ab initio} theory, isotopic studies, and
rotational structure. In this way a molecular carrier at 443\,nm was
very recently identified as C$_5$H$_5$ \cite{araki04}.

It has been suggested that some of the RRBs are due to
$\tilde{a}^3\Pi_u\rightarrow\tilde{X}^1\Sigma^+_g$ phosphorescence
\cite{Glinski97} of C$_3$ (which has never been observed in the gas
phase). Indeed, CO is seen in the Red Rectangle due to its
phosphorescence: the so-called Cameron Bands (between 1850 and 2600
\AA). If the species corresponding to the molecular carrier of the
RRBs is produced in a hydrocarbon discharge, but the emission
observed is due to phosphorescence, then it is likely that the
molecules will pass out of the light collection region and into the
vacuum pump before emitting a detectable number of photons
(phosphorescence typically occurs on the ms time-scale, and
molecular beams move at about 1mm/$\mu$s). An experiment designed to
circumvent this problem is described in section \ref{constr}.

In Fig. \ref{rrspec}, the observed spectrum of one of the RRBs is
displayed alongside a simulated spectrum performed by the authors.
The seemingly convincing simulation was performed with
$A''=0.84\,\textrm{cm}^{-1}$ and $A'=0.76\,\textrm{cm}^{-1}$. Such a
change of rotational constant (10\%) upon excitation is unusual yet
not unheard of. One class of molecule with $A$ constants very
sensitive to excited state are the carbenes. These molecules possess
lone-pair electrons which, when excited, bring about large changes
in geometry and thus rotational constant \cite{schmidt99}. A
rotational constant in the range given is slightly unusual. It is
too large to be due to three collinear second period atoms, so must
be accounted for by an effective diatomic (or some other slightly
non-collinear structure \cite{araki03}). Candidates include radical
molecules such as those observed in discharges by R2C2PI
spectroscopy \cite{ding03c,schmidt03} and rotational spectroscopy
\cite{travers97}. Of these, C$_7$H$_3$ \cite{ding03c} has a
structure which has calculated rotational constants in the ground
state consistent with the observed spectrum.

\section{Under construction: where to from here?} \label{constr}

The identification of extraterrestrial molecules in the optical
region can occur in two ways. Either the spectrum is recorded
firstly in the laboratory and subsequently in an extraterrestrial
object, or the absorption or emission line is observed by astronomy
and subsequently in the laboratory. Neither approach has been
particularly successful. A search for C$_5$ \cite{maier04}, which we
know to exist in the ISM \cite{bernath89}, at optical wavelengths,
was unsuccessful. It was concluded that the column density was only
one order of magnitude too low for optical detection. However, a
search for lines which might match C$_{60}^+$ turned up two
promising features \cite{foing94}. Unfortunately, the gas-phase
optical spectrum of C$_{60}^+$ is unknown and thus this
identification required confirmation. As described above, there are
hundreds of unidentified absorption and emission features in
astronomical spectra. Identifying these is a job for laboratory
spectroscopy. Many avenues have been explored, including a host of
carbon chain species and PAH cations. New experiments, presently
under construction, are described below.

\subsection{Spectroscopy of exotic cations}

Cation spectroscopy is difficult. They possess a much higher density
of states than neutral species and as such often have efficient
internal conversion pathways. As a consequence, only small cations
fluoresce. LIF is thus of limited applicability. Direct absorption
measurements by CRDS are possible. However, the species must have a
density in the free-jet expansion above a threshold limit for
detection. This technique is also mass-unresolved and thus
identification of a band carrier is often not straightforward.
R2C2PI spectroscopy is currently only applied to neutral species,
but in principle could be applied to cations if the mass-to-charge
ratio can be changed in a resonant process. Since a single ion can
be detected, this technique does not suffer from the problems of
sensitivity which plagues CRDS. Cations are difficult to doubly
ionize. As they are already charged, removal of an extra electron is
approximately twice as difficult as the first. The most amenable
example may be C$_{60}^+$. The ionization potential of C$_{60}$ is
7.62\,eV, and that of C$_{60}^+$ is 11.4\,eV
(109\,nm)\cite{steger92}. These photon energies required are only
just becoming convenient. Another problem is the internal conversion
mechanisms which preclude LIF from being applied to cations. A
consequence of internal conversion is that the double ionization
step must occur from the electronic ground state. Signal will be
very sensitive to the photon energy of the second laser pulse. One
unexplored direction is the implementation of ultrafast lasers for
ionization ($\tau_{FWHM}\approx100\textrm{fs}$). The ionization
laser need not be high-resolution (and necessarily are not due to
the time-energy uncertainty principle). One advantage of ultrafast
laser pulses is that their wavelength can be changed by non-linear
optical techniques with high efficiency (due to the high
peak-power). Thus, deep UV wavelengths may be accessed more easily
than with nanosecond laser pulses. The ionization step may also be
effected by multiphoton processes, which might be called resonant
2-colour multi-photon double-ionization spectroscopy (R2CMPDI).

One problem of C$_{60}^+$, and for large cations in general, is that
they are difficult to place into the gas phase and they are
difficult to cool to temperatures comparable to the interstellar
medium. One solution is to trap the ions and cool them with a buffer
gas \cite{schlemmer02}. In this way, it is possible to load an ion
trap with mass-selected C$_{60}^+$ or another large cationic species
and then cool to 5\,K or higher with a helium buffer. Spectroscopy
is then performed in the trap. This may be done either by R2CMPDI,
or by REMPD as in Ref. \cite{rolland03}.

Mass selection prior to trapping opens up the possibility of
performing spectroscopy on derivatives of PAH cations. It has been
observed by one of the authors (TWS), that nascent hydrocarbon
cations produced in a R2C2PI experiment will readily absorb photons
of energy $\approx6$\,eV and shed hydrogen atoms. In this way, the
signal observed for C$_9$H$_3^+$ also yielded the same resonance
enhanced ion signal at the masses for C$_9$H$_2^+$, C$_9$H$^+$ and
C$_9^+$ \cite{schmidt03}. It is thus likely that, in the
interstellar medium, PAH cations will absorb UV photons and shed
hydrogen atoms. This process will be in equilibrium with a hydrogen
capture process (ion-atom reaction) and it is possible that the
derivativized population of PAH cations and neutrals will be
significant. It is worth noting that a mono-dehydrogenated PAH
neutral will have a $\pi$-electronic structure similar to its cation
and will likely absorb in the visible. Obvious candidates for these
studies are naphthalenylium and anthracenylium cations and the
corresponding neutrals.

\subsection{Phosphorescence spectroscopy}

Fluorescence spectroscopy is performed by observing the emission, by
molecules, of photons at the point of laser-molecule interaction.
Where the emission lifetime is much longer than ns-$\mu$s, the
emission cannot be observed in this way. For this reason, relatively
little is known of laser-induced phosphorescence (LIP) spectroscopy,
and indeed about forbidden transitions in exotic molecules.

It has been speculated that the RRBs may be due to phosphorescence
\cite{Glinski97}. While it is unlikely that forbidden transitions
play any part in the DIBs, if a molecule is formed in a triplet
state by some reactive mechanism, the radiative lifetime is
irrelevant: molecules experience long delays between collisions in
the rarified environments of molecular clouds and nebulae (10 to
10,000\,s).

One way to observe phosphorescence in the laboratory is to excite a
forbidden transition in the gas-phase with a powerful laser, then
freeze the triplet excited molecules onto a substrate at 5\,K. The
ensuing phosphorescence is then detected at leisure as the molecules
are now frozen into a matrix of the carrier gas (e.g. argon).
Phosphorescence as a function of laser wavelength yields a gas-phase
phosphorescence spectrum which may be compared to astronomical
spectra. This technique has already been applied to the spectroscopy
of benzaldehyde \cite{lip}.

\section{Concluding Remarks}

Models of interstellar chemistry are not only tested by predicting
observations of column densities by millimetre-wave spectroscopy,
but must also predict abundances of species without permanent dipole
moments to which millimetre-wave spectroscopy is blind. The
identification of molecules in the optical region of the
electromagnetic spectrum requires high resolution astronomical
observation coupled with sophisticated laboratory experiments. While
astronomical observations have uncovered hundreds of unidentified
and presumably molecular absorption and emission features, and
laboratory spectroscopy has produced cold, gas-phase spectra of
hundreds of candidate carriers, there is as of yet not one certain
match between a DIB or RRB and a laboratory spectrum. C$_3$ has been
observed in molecular clouds, but so far C$_5$ has been elusive in
the optical region.

As far as identification of the DIBs is concerned, the outstanding
candidates within the frame of the carbon chain hypothesis are the
odd numbered pure carbon clusters. These chains possess strong
transitions which increase linearly with the size of the chain. They
are also known to absorb in the visible region. Other candidates yet
to be tested are the iso-electronic monohydrogenated carbon chain
cations. These are only now being studied in the condensed phase.
Gas-phase spectroscopy of cations is a field under development. The
coming years should see some progress. An outstanding yet unobserved
spectrum is that of C$_{60}^+$ in the gas-phase. This spectrum, when
obtained, will confirm whether or not this cation is abundant in the
interstellar medium. Modeling its formation should be a great
challenge for theoretical astrochemists.

The spectroscopy of PAHs in the laboratory is ongoing business. Very
few gas-phase spectra of PAH cations and derivatives have been
observed. More studies are needed before an informed opinion can be
formed on the importance of PAHs with respect to the DIBs. One of
the great challenges facing these studies is the methodology with
which the spectra of cations may be observed in the gas-phase.

The RRBs remain unidentified. It is clear that the carrier emits,
and as such it will be observed in the laboratory by observation of
its laser-induced emission. The challenge to the experimentalist is
to build the laser-induced fluorescence/phosphorescence apparatus
and find a way of making the presumably exotic carrier \textit{in
situ}. The carrier must be abundant in the Red Rectangle nebula and
thus should be produced in a discharge of the right precursor
mixture. Whether the carrier possesses a heteroatom remains to be
seen.

The identification of molecules in the interstellar medium is an
on-going quest. The coming decade should see the identification of
several of the DIBs, or if not, then certainly the gas-phase spectra
of troublesome cations will be obtained. On this quest, physical
chemists and astronomers walk together in an example of cooperation
and collaboration between two seemingly different fields of
scientific endeavour.

\end{document}